\documentclass[aps,prl,twocolumn,superscriptaddress,showpacs,preprintnumbers,amsmath,amssymb]{revtex4}
\usepackage{amsmath} 
\usepackage{graphicx} 
\usepackage{dcolumn}  
\usepackage{soul}
\usepackage{color}
\usepackage{graphicx}

\graphicspath{{ps}}

\begin{document}


\preprint{\vbox{ \hbox{   }
                               \hbox{Belle Preprint 2010-18}
                               \hbox{KEK Preprint 2010-29}
}}

\title{ \quad\\[1.0cm] 
Search for $CP$-violating charge asymmetry in $B^{\pm} \to J/\psi K^{\pm}$ decays}

\affiliation{Budker Institute of Nuclear Physics, Novosibirsk}
\affiliation{Faculty of Mathematics and Physics, Charles University, Prague}
\affiliation{Chiba University, Chiba}
\affiliation{University of Cincinnati, Cincinnati, Ohio 45221}
\affiliation{Department of Physics, Fu Jen Catholic University, Taipei}
\affiliation{Justus-Liebig-Universit\"at Gie\ss{}en, Gie\ss{}en}
\affiliation{The Graduate University for Advanced Studies, Hayama}
\affiliation{Hanyang University, Seoul}
\affiliation{University of Hawaii, Honolulu, Hawaii 96822}
\affiliation{High Energy Accelerator Research Organization (KEK), Tsukuba}
\affiliation{Indian Institute of Technology Guwahati, Guwahati}
\affiliation{Institute of High Energy Physics, Chinese Academy of Sciences, Beijing}
\affiliation{Institute of High Energy Physics, Vienna}
\affiliation{Institute of High Energy Physics, Protvino}
\affiliation{Institute for Theoretical and Experimental Physics, Moscow}
\affiliation{J. Stefan Institute, Ljubljana}
\affiliation{Kanagawa University, Yokohama}
\affiliation{Institut f\"ur Experimentelle Kernphysik, Karlsruher Institut f\"ur Technologie, Karlsruhe}
\affiliation{Korea Institute of Science and Technology Information, Daejeon}
\affiliation{Korea University, Seoul}
\affiliation{Kyungpook National University, Taegu}
\affiliation{\'Ecole Polytechnique F\'ed\'erale de Lausanne (EPFL), Lausanne}
\affiliation{Faculty of Mathematics and Physics, University of Ljubljana, Ljubljana}
\affiliation{University of Maribor, Maribor}
\affiliation{Max-Planck-Institut f\"ur Physik, M\"unchen}
\affiliation{University of Melbourne, School of Physics, Victoria 3010}
\affiliation{Nagoya University, Nagoya}
\affiliation{Nara Women's University, Nara}
\affiliation{National Central University, Chung-li}
\affiliation{National United University, Miao Li}
\affiliation{Department of Physics, National Taiwan University, Taipei}
\affiliation{H. Niewodniczanski Institute of Nuclear Physics, Krakow}
\affiliation{Nippon Dental University, Niigata}
\affiliation{Niigata University, Niigata}
\affiliation{University of Nova Gorica, Nova Gorica}
\affiliation{Novosibirsk State University, Novosibirsk}
\affiliation{Osaka City University, Osaka}
\affiliation{Panjab University, Chandigarh}
\affiliation{University of Science and Technology of China, Hefei}
\affiliation{Seoul National University, Seoul}
\affiliation{Sungkyunkwan University, Suwon}
\affiliation{School of Physics, University of Sydney, NSW 2006}
\affiliation{Tata Institute of Fundamental Research, Mumbai}
\affiliation{Excellence Cluster Universe, Technische Universit\"at M\"unchen, Garching}
\affiliation{Toho University, Funabashi}
\affiliation{Tohoku Gakuin University, Tagajo}
\affiliation{Tohoku University, Sendai}
\affiliation{Department of Physics, University of Tokyo, Tokyo}
\affiliation{Tokyo Metropolitan University, Tokyo}
\affiliation{Tokyo University of Agriculture and Technology, Tokyo}
\affiliation{IPNAS, Virginia Polytechnic Institute and State University, Blacksburg, Virginia 24061}
\affiliation{Wayne State University, Detroit, Michigan 48202}
\affiliation{Yonsei University, Seoul}
  \author{K.~Sakai}\affiliation{Niigata University, Niigata} 
  \author{T.~Kawasaki}\affiliation{Niigata University, Niigata} 
  \author{H.~Aihara}\affiliation{Department of Physics, University of Tokyo, Tokyo} 
  \author{K.~Arinstein}\affiliation{Budker Institute of Nuclear Physics, Novosibirsk}\affiliation{Novosibirsk State University, Novosibirsk} 
  \author{T.~Aushev}\affiliation{\'Ecole Polytechnique F\'ed\'erale de Lausanne (EPFL), Lausanne}\affiliation{Institute for Theoretical and Experimental Physics, Moscow} 
  \author{A.~M.~Bakich}\affiliation{School of Physics, University of Sydney, NSW 2006} 
  \author{V.~Balagura}\affiliation{Institute for Theoretical and Experimental Physics, Moscow} 
  \author{E.~Barberio}\affiliation{University of Melbourne, School of Physics, Victoria 3010} 
  \author{K.~Belous}\affiliation{Institute of High Energy Physics, Protvino} 
  \author{V.~Bhardwaj}\affiliation{Panjab University, Chandigarh} 
% ...
\collaboration{The Belle Collaboration}


\begin{abstract}

We present the result of a search for
charge asymmetry in $B^{\pm} \to J/\psi K^{\pm}$ decays
using 772 $\times$ 10$^6$ $B\overline{B}$
pairs collected at the $\Upsilon(4S)$
 resonance by the Belle detector at the KEKB asymmetric-energy $e^+ e^-$ collider.
The $CP$-violating charge asymmetry is measured to be $A_{CP}(B^+ \to J/\psi K^+) = [-0.76 \pm 0.50~\rm(stat) \pm 0.22~\rm(syst)]\%$.
\end{abstract}


\maketitle

\tighten

{\renewcommand{\thefootnote}{\fnsymbol{footnote}}}
\setcounter{footnote}{0}

 Violation of $CP$ symmetry in the Standard Model (SM) has been well established.
Interest has now shifted to the search for new sources of $CP$ violation due to physics beyond the SM,
since the $CP$ violation content of the SM does not explain
the matter-antimatter asymmetry of the Universe~\cite{CP}.
In the SM, $CP$-violating
phenomena in the quark sector are described by the Kobayashi-Maskawa theory~\cite{KM},
in which a single irreducible complex phase gives rise to all $CP$-violating asymmetries.
The decay $B^+ \to J/\psi K^+$ is mediated by a color-suppressed 
$\bar{b} \to \bar{c} c \bar{s}$ transition, where the dominant $\bar{b} \to \bar{c}$ tree-level 
amplitude and the $\bar{b} \to \bar{s}$ penguin amplitude have a small relative complex phase,
arg $[-V_{cs}V_{cb}^* / V_{ts}V_{tb}^* ]$~\cite{KM} in the SM~\cite{CC}.
The charge asymmetry in the decay $B^+ \to J/\psi K^+$ is defined as
\begin{equation}
\begin{split}
A_{CP}&(B^+ \to J/\psi K^+) \\ \nonumber
  &= \frac{\mathcal{B}(B^- \to J/\psi K^-) - \mathcal{B}(B^+ \to J/\psi K^+)}
        {\mathcal{B}(B^- \to J/\psi K^-) + \mathcal{B}(B^+ \to J/\psi K^+)},
\label{equ:acp_difinition}
\end{split}
\end{equation}

\noindent
where $\mathcal{B}$ denotes the branching fraction.
Direct $CP$ violation would appear as a nonzero $A_{CP}(B^+ \to J/\psi K^+)$ value
and is predicted to be quite small, $\simeq$ 0.3\%~\cite{REF1}, in the SM.
Some new physics models predict enhanced values of this asymmetry~\cite{REF1}.
For example, a model with an extra $U(1)'$ gauge boson~\cite{REF2} and
another model with an extra coupling to the charged Higgs boson~\cite{REF3}
predict asymmetries of $\mathcal{O}$(1\%) and $\mathcal{O}$(10\%), respectively.

The $B^+ \to J/\psi K^+$ decay mode has low backgrounds and
a large branching fraction, when compared to that of other charmonium decay modes.
These properties enable a precise $A_{CP}(B^+ \to J/\psi K^+)$ measurement, 
which provides an important test of various new physics models
and constrains their parameter spaces.

The current world average for $A_{CP}(B^+ \to J/\psi K^+)$ 
is (0.9 $\pm$ 0.8)\%~\cite{PDG} which is dominated by the D0 result, (0.75 $\pm$ 0.61 $\pm$ 0.30)\%~\cite{D0}, while the most precise result from an $e^+e^-$ collider experiment
is the {\it BABAR} result of (3.0 $\pm$ 1.4 $\pm$ 1.9)\%~\cite{BABAR}.

In this paper, we report a measurement of $A_{CP}(B^+ \to J/\psi K^+)$ 
using a 711 fb$^{-1}$ data set that contains 772 $\times$ $10^6$ $B\overline{B}$ pairs.
The $B^+ \to J/\psi K^+$ decay is reconstructed in the $J/\psi \to \ell^+\ell^-$
($\ell = e$ or $\mu$) channels.  For a precise measurement of the charge 
asymmetry in this decay, the asymmetry of charged kaon detection 
efficiencies must be carefully studied and corrected for.
The asymmetry in detection efficiency arises due to 
the asymmetric geometry of the detector, different interaction rates of kaons in the detector
material, and differences in kaon identification efficiencies for $K^+$ and $K^-$.

KEKB is an asymmetric electron-positron storage ring that collides 8.0 GeV electrons with 3.5 GeV positrons at the $\Upsilon(4S)$ resonance
(center-of-mass [c.m.] energy $\sqrt{s} = 2E_{\rm{beam}} = 10.58$ GeV).
The $\Upsilon(4S)$ resonance is boosted by $\beta\gamma = 0.425$~\cite{KEKB}.
The Belle detector is a large-solid-angle magnetic
spectrometer that consists of a silicon vertex detector (SVD),
a 50-layer central drift chamber (CDC), an array of
aerogel threshold Cherenkov counters (ACC),
a barrel-like arrangement of time-of-flight
scintillation counters (TOF), and an electromagnetic calorimeter (ECL)
comprised
CsI (Tl) crystals located inside a superconducting
solenoid coil that provides a 1.5~T
magnetic field. An iron flux-return located outside of
the coil is instrumented to detect $K_L^0$ mesons and to identify
muons (KLM). The detector is described in detail elsewhere~\cite{BELLE-DETECTOR, svd2}.

Hadronic events are initially
selected by requiring at least three reconstructed charged tracks,
a total reconstructed ECL energy in the c.m.~in the range
0.1$\sqrt{s}$ $-$ 0.8$\sqrt{s}$,
at least one large-angle cluster in the ECL,
a total visible energy -- calculated from
all charged tracks and isolated neutral showers -- greater than
0.2$\sqrt{s}$,
an absolute value of the $z$ component of the c.m.~momentum less than
0.5$\sqrt{s}$,
and a reconstructed primary vertex that is consistent with the known location of the interaction point.
To suppress two-jet non-$\Upsilon(4S)$ background relative to $B\overline{B}$ events, we require 
$R_2 < 0.5$, where $R_2$ is the ratio of the second to zeroth Fox-Wolfram 
moments~\cite{FW}.
To remove charged particle tracks that are poorly measured or do not come from the interaction region,
we require $dz < 5$ cm for all tracks, where $dz$ is the absolute value of the coordinate along
the beam direction at the point on the track closest to the origin.

The $J/\psi$ meson is reconstructed from one tightly and one loosely
identified lepton.
For muon tracks, the tight identification criterion is based on track penetration depth and hit scatter in the KLM system,
while the loose identification criterion requires only that the
tracks have an energy deposit in the ECL that 
is consistent with that of a minimum ionizing particle.
Electron tracks are tightly identified by a requirement on a combination
of $dE/dx$ from the CDC, $E/p$ ($E$ is the energy deposit in the ECL and
the $p$ is momentum measured by the SVD and the CDC), and shower shape in the ECL.
For loose identification, either $dE/dx$ or $E/p$ is required to be consistent with the electron hypothesis.
We correct for final-state radiation or bremsstrahlung in
the inner parts of the detector by including in the $e^+e^-$ invariant mass calculation
the four-momentum of every photon detected within 0.05 rad of the original electron or positron direction.
Since small residual radiative tails still remain,
we use asymmetric invariant mass requirements,
$-60$ MeV/$c^2 < M_{\mu^+\mu^-}-m_{J/\psi} < 36$ MeV/$c^2$ and  
$-150$ MeV/$c^2 < M_{e^+e^-(\gamma)}-m_{J/\psi} < 36$ MeV/$c^2$, for the $\mu^+\mu^-$ and $e^+e^-$ pairs, respectively.

The combined information from the CDC , TOF, and ACC is used to form
a $K-\pi$ likelihood ratio,
$\mathcal{R}_K = \mathcal{L}_K / (\mathcal{L}_K + \mathcal{L}_{\pi})$, where $\mathcal{L}_{K(\pi)}$ is the likelihood of
the kaon (pion) hypothesis.
We require $\mathcal{R}_K > 0.6$ for kaon candidates,
which is approximately 80\% efficient for kaons, while giving a
misidentification probability of below 10\% for pions.

Charged $B$ mesons are reconstructed by combining a $J/\psi$
candidate with a charged kaon candidate.
The energy difference, $\Delta E \equiv E_{\rm {cand}} - E_{\rm {beam}}$, and the beam-energy constrained mass,
$M_{\rm {bc}} \equiv \sqrt{E_{\rm {beam}}^2 - {\boldsymbol p}_{\rm {cand}}^2}$, are used to separate signal from background
[$E_{\rm {cand}}$ and ${\boldsymbol p}_{\rm {cand}}$ are the $B$ candidate energy and momentum, calculated in the $\Upsilon(4S)$ c.m.~system, after performing a
mass- and vertex-constrained fit to the leptons from the $J/\psi$ decay].

\begin{figure}
  \includegraphics[scale=0.4]{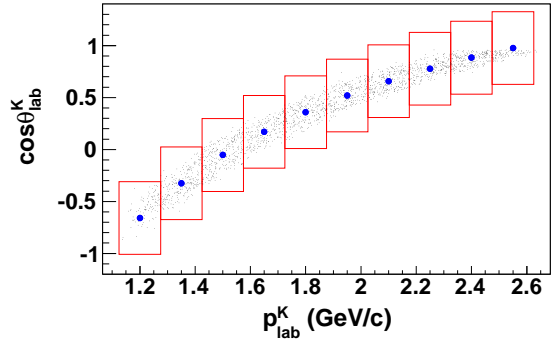}
  \caption{(color online). 
 Illustration of the ($p_{\rm lab}^{K}, \cos\theta_{\rm lab}^{K}$) binning.
 $p_{\rm lab}^{K}$ is divided into ten bins each with a width of
0.15 GeV/$c$.
 The $\cos\theta_{\rm lab}^{K}$ bin width is
0.7, and fully contains the $B^+ \to J/\psi K^+$ signal.
 The bins are centered at the kinematically determined
 values for the two-body decay (blue circles).
The small black dots in the figure are $B^+ \to J/\psi K^+$ events generated by a Monte Carlo simulation.
 }
  \label{fig:binning}
\end{figure}

\begin{figure}
  \includegraphics[scale=0.4]{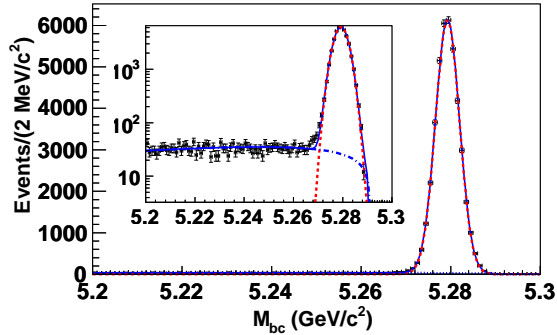}
  \caption{(color online). 
  $M_{\rm bc}$ distribution of $B^+ \to J/\psi K^+$ candidates
  summed over all bins and for both
$B$ charge states (inset plot is on a semilog scale).
The blue solid, blue dot-dashed, and red dashed curves are
the total fit, the background and the signal components, respectively.
}
  \label{fig:b_all}
\end{figure}

In order to determine the signal yield and charge asymmetry, we fit
$M_{\rm {bc}}$ distributions after requiring $|\Delta E| <$ 40 MeV.
Each $M_{\rm {bc}}$ distribution for 5.2 GeV/$c^2 < M_{\rm {bc}} < 5.3$ GeV/$c^2$ is
fitted with the sum of a Gaussian for the signal and
an ARGUS function~\cite{ARGUS} for background.
We simultaneously fit the $B^+$ and $B^-$ distributions with common shape 
parameters for the signal and background, except for the mean of
the signal Gaussians.  In the fit we assume that there is no asymmetry in the background; the effect of
a possible asymmetry is included in the systematic error evaluation.
We neglect a small contribution from correlated $B$ background,
which peaks at the signal position in the $M_{\rm bc}$ distribution.
The effect of this "peaking background" is included in the systematic error as described below.
Shape parameters and normalizations are allowed to vary in the fit. 
To correct for the kaon detection asymmetry, which depends on the
momentum $p_{\rm lab}^{K}$ and polar angle $\cos\theta_{\rm lab}^{K}$ 
of kaons in the laboratory system, we perform fits in ten bins of  
($p_{\rm lab}^{K}, \cos\theta_{\rm lab}^{K}$).
The binning is shown in Fig.~\ref{fig:binning}.
The bins are labeled 1 to 10 with
increasing $p_{\rm lab}^{K}$.  We observe
a total signal yield of 41,315 $\pm$ 205 events.
The $M_{\rm bc}$ distribution for all bins combined is shown in Fig.~\ref{fig:b_all}.  
The measured raw asymmetry ($A_{CP}^{\rm raw}$) in each bin is given in
Table \ref{tbl:summary}.

We measure the kaon detection asymmetry, $A_{\varepsilon}^{K^+}$, in data using 
$D_s^+ \to \phi\pi^+$ ($\phi \to K^+K^-$)
and $D^0 \to K^-\pi^+$ decay modes~\cite{BRKO}.
Here, we denote $A^{x^+} \equiv \frac{N^{x^+} - N^{x^-}}{N^{x^+} + N^{x^-}}$.
The measured asymmetries of the above modes can be written as
\begin{equation}
\begin{split}
A_{\rm rec}^{D_s^+} &= A_{\rm FB}^{D_s^+} + A_{\varepsilon}^{\pi^+} \\ \nonumber
A_{\rm rec}^{D^0}   &= A_{\rm FB}^{D^0}   + A_{\varepsilon}^{\pi^+} - A_{\varepsilon}^{K^+},
\label{equ:asymmetry_of_d}
\end{split}
\end{equation}
assuming the asymmetries are small.
Here $A_{\rm {FB}}$ denotes the forward-backward asymmetry due to the $\gamma^* - Z^0$ interference in
 $e^+e^- \to c\bar{c}$
and $A_{\varepsilon}^{\pi^+}$ is the pion detection asymmetry.
We can extract $-A_{\varepsilon}^{K^+}$ by subtracting the 
$A_{\rm {rec}}^{D_s^+}$ value from the $A_{\rm {rec}}^{D^0}$ with the assumption 
$ A_{\rm FB}^{D_s^+} = A_{\rm FB}^{D^0}$.
This assumption is reasonable
because the effect of the fragmentation on the forward-backward asymmetry is expected to be small.
Possible deviations from this assumption are checked in data and included as a small contribution to the systematic error.
The subtraction is performed in bins of pion momentum, $p_{\rm {lab}}^{\pi}$,
and polar angle in the laboratory system, $\cos\theta_{\rm {lab}}^{\pi}$,
and the charmed meson's polar angle in the c.m.~system,
$\cos\theta_{\rm {CMS}}^{D}$ (since $\cos\theta_{\rm {CMS}}^{D}$ is
correlated with $\cos\theta_{\rm {lab}}^{\pi}$ and $A_{\rm FB}^{D_s^+}$
depends on it).

\begin{figure}
  \includegraphics[scale=0.35]{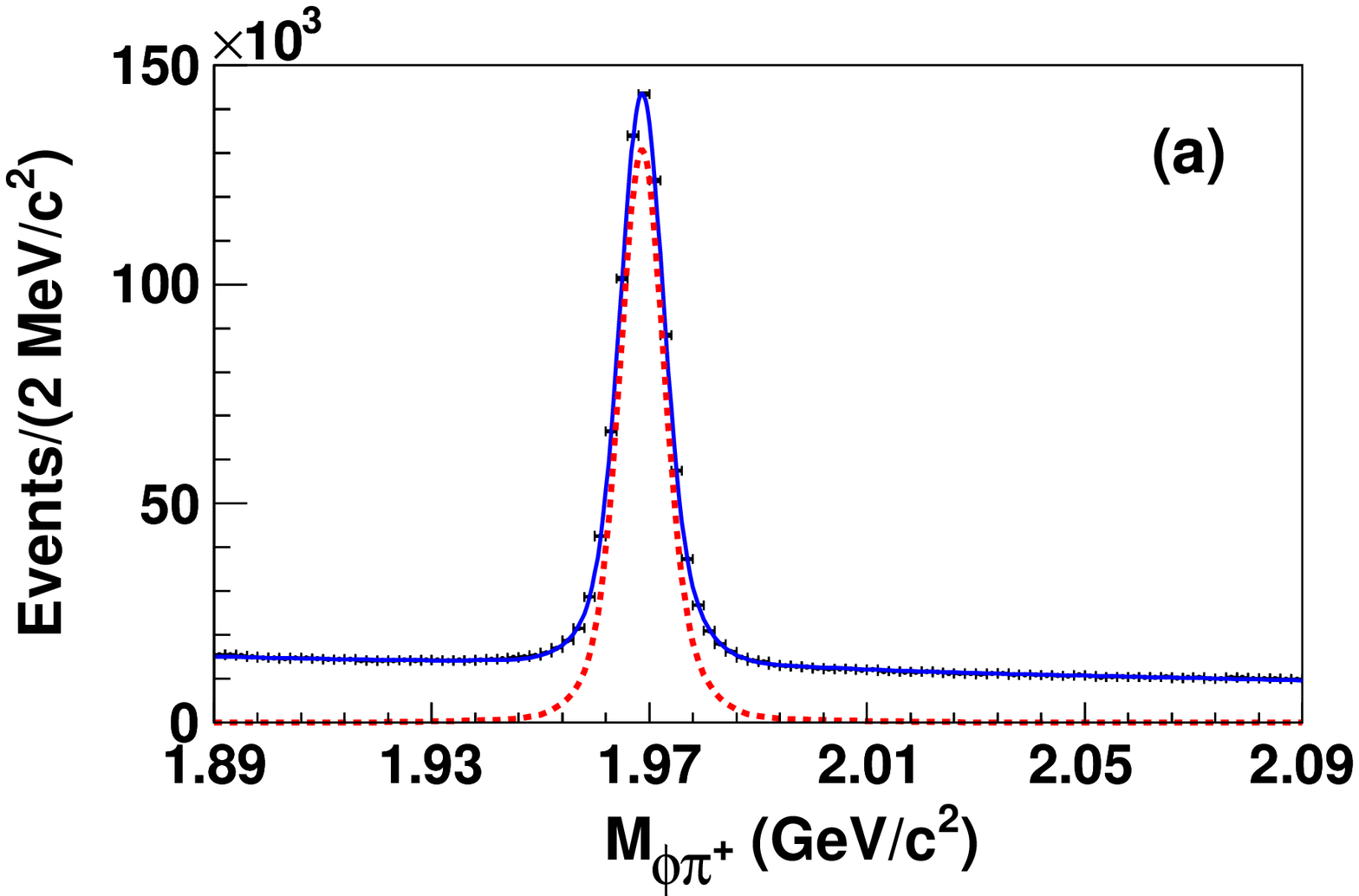}
  \includegraphics[scale=0.35]{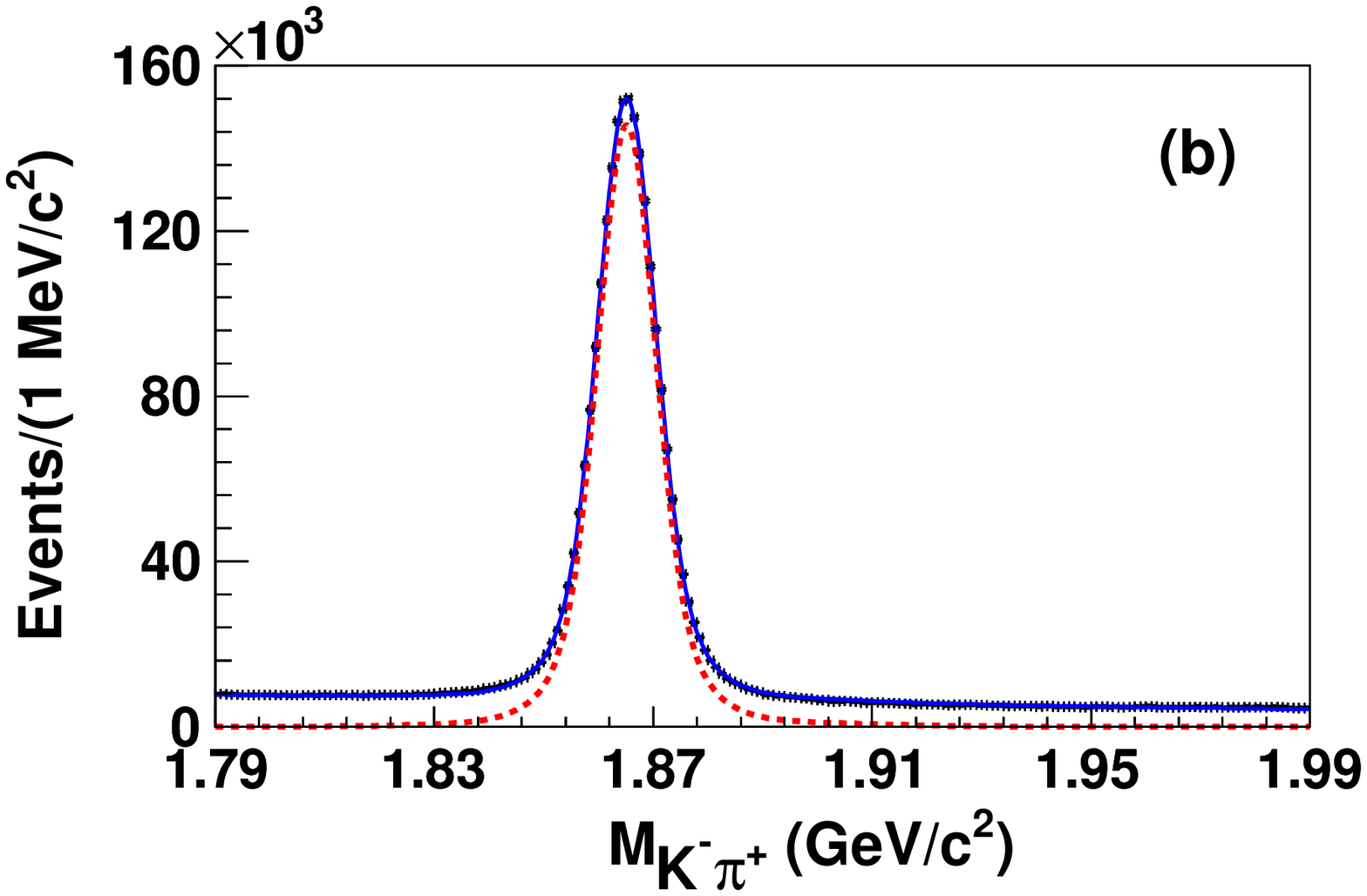}
  \caption{
(color online). $D_s^+ \to \phi \pi^+$ (a) and $D^0 \to K^-\pi^+$ (b) invariant mass distribution
  summed over all bins including 
the charge-conjugate final states.
The blue curve shows the results of the fit described in text, and the red dashed
curve shows the signal.
}
  \label{fig:d_all}
\end{figure}

We reconstruct $D_s^+ \to \phi\pi^+$ and $D^0 \to K^-\pi^+$ decays
with charged tracks that originate from the vicinity of the interaction point.
We require $\mathcal{R}_K > 0.6$ for kaons and $\mathcal{R}_K < 0.4$ for pions.
The $\phi$ meson candidates are selected with the requirement
1.00 GeV/$c^2 < M_{K^+K^-} <$ 1.04 GeV/$c^2$.
To remove $D_s^+$ and $D^0$ mesons produced in $B$ meson decays, we require the charmed meson momentum 
in the c.m.~system be greater than 2.5 GeV/$c$.
The invariant mass distributions of $D_s^+ \to \phi\pi^+$ and 
$D^0 \to K^-\pi^+$ candidates after these requirements are shown in Figs.~\ref{fig:d_all}.

\begin{figure}
  \includegraphics[scale=0.45]{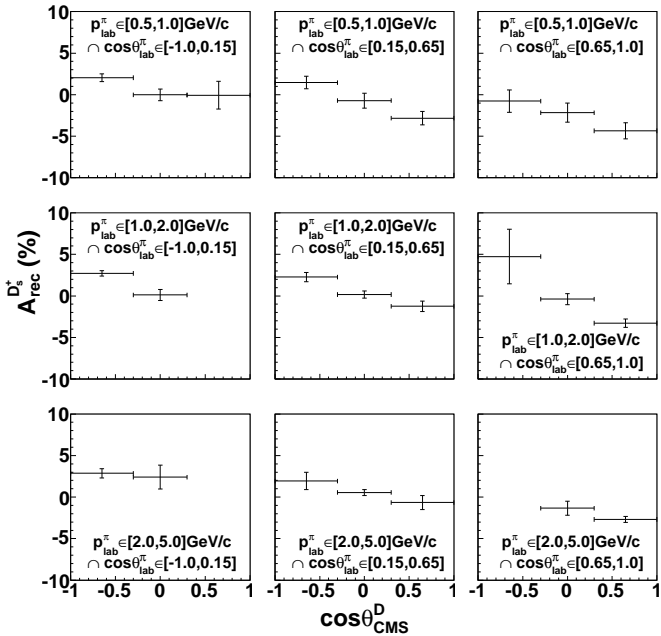}
  \caption{
 $A_{\rm rec}^{D_s^+}$ map in bins of 
 ($p_{\rm {lab}}^{\pi}$, $\cos\theta_{\rm {lab}}^{\pi}$,
 $\cos\theta_{\rm {CMS}}^{D}$).
The errors shown here are the statistical errors in
the $D_s^+ \to \phi \pi^+$ signal yield.
The empty bins do not contain enough $D_s^+$ candidates to obtain $A_{\rm rec}^{D_s^+}$ and hence we assign the value of 0.0\%.
  }
  \label{fig:ads_results}
\end{figure}

We first obtain the $A_{\rm rec}^{D_s^+}$ map in three-dimensional (3D)
bins of ($p_{\rm {lab}}^{\pi}$, $\cos\theta_{\rm {lab}}^{\pi}$,
$\cos\theta_{\rm {CMS}}^{D}$).  
In each bin, $A_{\rm rec}^{D_s^+}$ is obtained by fitting
the reconstructed $D_s^+$ candidate mass distributions
(1.89 GeV/$c^2 < M_{\phi\pi^+} < 2.09$ GeV/$c^2$).  The fits
are performed in a manner similar to those performed for $B^+ \to J/\psi K^+$.
$D_s^+$ signals are parameterized as a sum of two Gaussian
functions and a bifurcated Gaussian, which represents 
the tail of the distribution
(for high-statistics bins only). The bin-dependent fractions of Gaussians
are fixed to those obtained from Monte Carlo simulations.
The background is parametrized as a first-order polynomial and
its asymmetry is allowed to vary in the fits.
Figure~\ref{fig:ads_results} shows the $A_{\rm rec}^{D_s^+}$ map in
bins of ($p_{\rm {lab}}^{\pi}$, $\cos\theta_{\rm {lab}}^{\pi}$,
$\cos\theta_{\rm {CMS}}^{D}$).  The 3D binning ($3\times 3\times 3$)
is selected to have sufficient granularity with large enough statistics
in each bin.

\begin{figure}
  \includegraphics[scale=0.4]{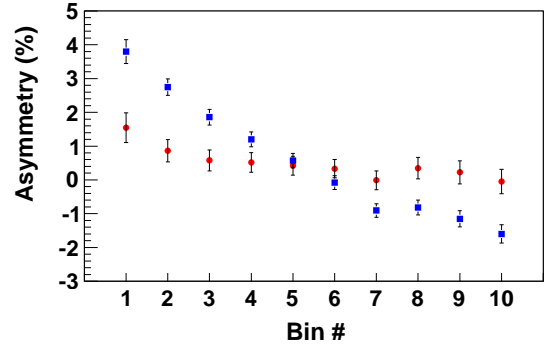}
  \caption{
(color online).
$A_{\rm{rec}}^{D^0}$ (blue squares) and corrected $A_{\rm{rec}}^{D^0}$ (red circles), which corresponds to $-A_{\rm{\varepsilon}}^{K^+}$ in ($p_{\rm lab}^{K}, \cos\theta_{\rm lab}^{K}$) bins.
The errors in the corrected values of $A_{\rm{rec}}^{D^0}$ include the statistical errors of the $A_{\rm rec}^{D_s^+}$.
}
  \label{fig:compare_a_d}
\end{figure}

The $A_{\varepsilon}^{K^+}$ values are extracted using 
$D^0 \to K^-\pi^+$ decays as follows: fits are performed to the $D^0$
candidate invariant mass distribution
in 1.79 GeV/$c^2 < M_{K^-\pi^+} <$ 1.99 GeV/$c^2$
with a parameterization similar to that used for the $D_s^+$.
Here $N_{\rm{rec}}^{D^0}$ and $N_{\rm{rec}}^{\bar{D^0}}$ are corrected 
according to the $A_{\rm rec}^{D_s^+}$ values in bins of
($p_{\rm {lab}}^{\pi}$, $\cos\theta_{\rm {lab}}^{\pi}$, $\cos\theta_{\rm {CMS}}^{D}$). 
The obtained values of $N_{\rm{rec}}^{D^0}$ and $N_{\rm{rec}}^{\bar{D^0}}$ are already corrected for 
$A_{\rm FB}$ and $A_{\varepsilon}^{\pi^+}$ and their asymmetry gives $-A_{\varepsilon}^{K^+}$.
The values of the kaon asymmetry $A_{\varepsilon}^{K^+}$ are determined for different
($p_{\rm lab}^{K}, \cos\theta_{\rm lab}^{K}$) bins defined in 
Fig.~\ref{fig:binning}.
These results are shown in Fig.~\ref{fig:compare_a_d} and the obtained kaon
asymmetry values are quoted in the last column of Table~\ref{tbl:summary}.

\begin{figure}
  \includegraphics[scale=0.4]{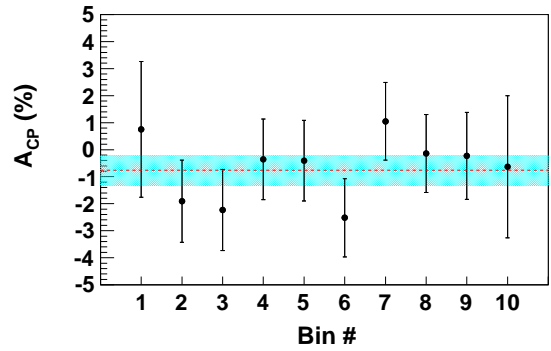}
  \caption{
    (color online). $A_{CP}$ distribution in
different bins.
Errors are obtained from $A_{CP}^{\rm raw}$ and $A_{\varepsilon}^{K^+}$ uncertainties summed in quadrature.
    The red dashed line shows the weighted average of $A_{CP}$ and its
uncertainty is shown as the light-blue band.
    The values obtained in the ten ($p_{\rm lab}^{K}, \cos\theta_{\rm lab}^{K}$) regions are consistent with each other.
   }
  \label{fig:acp_results}
\end{figure}

 Finally, the measured $A_{CP}^{\rm raw}$ values are corrected by $A_{\varepsilon}^{K^+}$
in each bin.
The results are shown
in Fig.~\ref{fig:acp_results} and summarized in Table~\ref{tbl:summary}.
We obtain $A_{CP}(B^+ \to J/\psi K^+) = (-0.76 \pm 0.50)\%$,
which is a weighted average of corrected asymmetries, and in which the error includes only the statistical error of the $A_{CP}^{\rm raw}$ determination.
The weights in the averaging procedure are also based only on the statistical error of the $A_{CP}^{\rm raw}$.
Statistical and systematic uncertainties of $A_{\varepsilon}^{K^+}$ are included in the systematic error as described below.

\begin{table}[htb]
\caption{
 Summary of asymmetries (in units of \%).
 $A_{CP}$ is the corrected charge asymmetry.
The final $A_{CP}$ values are weighted averages with the statistical errors
from $A_{CP}^{\rm raw}$.  The first error in
$A_{\varepsilon}^{K^+}$ is the statistical error of the $D^0 \to K^-\pi^+$ signal yield.
The second error comes from the statistical errors in
$A_{\rm{rec}}^{D_s^+}$ and has correlation among bins.
}
\label{tbl:summary}
\scalebox{0.85}{
\begin{tabular}
 {@{\hspace{0.5cm}}c@{\hspace{0.5cm}} @{\hspace{0.5cm}}c@{\hspace{0.5cm}} @{\hspace{0.5cm}}c@{\hspace{0.5cm}} @{\hspace{0.5cm}}c@{\hspace{0.5cm}}}
    \hline
    \hline
    Bin                              & $A_{CP}$                    & $A_{CP}^{\rm raw}$            & $A_{\varepsilon}^{K^+}$     \\
   \hline
    1                                & $+0.75$                           & $+2.30 \pm 2.47$                  & $-1.55 \pm 0.35 \pm 0.26$ \\
    2                                & $-1.91$                           & $-1.04 \pm 1.49$                  & $-0.86 \pm 0.24 \pm 0.22$ \\
    3                                & $-2.23$                           & $-1.65 \pm 1.47$                  & $-0.58 \pm 0.23 \pm 0.21$ \\
    4                                & $-0.36$                           & $+0.16 \pm 1.46$                  & $-0.52 \pm 0.22 \pm 0.19$ \\
    5                                & $-0.41$                           & $+0.01 \pm 1.46$                  & $-0.42 \pm 0.21 \pm 0.18$ \\
    6                                & $-2.52$                           & $-2.19 \pm 1.42$                  & $-0.33 \pm 0.20 \pm 0.18$ \\
    7                                & $+1.05$                           & $+1.04 \pm 1.41$                  & $+0.01 \pm 0.20 \pm 0.19$ \\
    8                                & $-0.14$                           & $+0.20 \pm 1.41$                  & $-0.35 \pm 0.22 \pm 0.23$ \\
    9                                & $-0.23$                           & $-0.01 \pm 1.57$                  & $-0.22 \pm 0.24 \pm 0.24$ \\
   10                                & $-0.63$                           & $-0.68 \pm 2.61$                  & $+0.05 \pm 0.27 \pm 0.24$ \\
    \hline
   Total            & $-0.76$                                            & $-0.33 \pm 0.50$                  & $-0.43 \pm 0.07 \pm 0.17$ \\
    \hline
    \hline
\end{tabular}
}
\end{table}

Systematic errors arise from three sources: the systematic uncertainty in 
$A_{CP}^{\rm raw}$ measurement, the uncertainty of the $A_{\varepsilon}^{K^+}$ due to
 $A_{\rm rec}^{D_s^+}$ and due to
$A_{\rm {rec}}^{D^0}$.
The systematic errors are summarized in Table~\ref{tbl:B_FITTING_RESULTS}.
The systematic uncertainties due to the choice of binning, fit range, and mass
windows are estimated from variations of the fit results,
obtained by refitting the data using different choices.

The dominant systematic error comes from the uncertainty
in $A_{\rm {rec}}^{D^+_s}$, 
in which the statistical error in 
$D_s^+ \to \phi \pi^+$ signal yields contributes 
0.17\%.  The choice of 3D binning in  ($p_{\rm {lab}}^{\pi}$, 
$\cos\theta_{\rm {lab}}^{\pi}$, $\cos\theta_{\rm {CMS}}^{D}$) contributes 0.08\%.
The kaon detection asymmetry in $\phi \to K^+K^-$ cancels if the
momentum distributions of $K^+$ and $K^-$ are identical in $D_s^+$
decay.  We find a small difference between them that
arises from asymmetry in the helicity angle distribution due to the
interference between the $\phi$ and $S$-wave component in 
$D_s^+ \to K^+K^-\pi^+$ decay.  We estimate the effect on our
measurement to be 0.05\% from the difference of momentum distributions
in data.
The effect of empty bins in the $A_{\rm {rec}}^{D^+_s}$ map is estimated
by setting $A_{\rm {rec}}^{D^+_s}$ values of empty bins to $\pm$2\% and
results in a negligibly small contribution of 0.001\%.
The uncertainty in $A_{\rm {rec}}^{D^0}$ mainly comes from the
statistical errors in the $D^0 \to K^-\pi^+$ (0.07\%)
and ($p_{\rm lab}^{K}, \cos\theta_{\rm lab}^{K}$) binning (0.04\%).
A possible $CP$ asymmetry in the $D^0 \to K^- \pi^+$ final state 
arises from the interference between decays with and without 
$D^0 - \bar{D^0}$ mixing.
The uncertainty is estimated from the 95\% confidence level upper limit 
on the $CP$-violating asymmetry, 
$A_{CP}^{D^0} = -y\sin\delta\sin\phi\sqrt{R}$~\cite{ACP-D0-13}, using the world
average of $D^0 - \bar{D^0}$ mixing and $CP$ violation 
parameters~\cite{ACP-D0-14} and is found to be 0.01\%.
$A_{CP}^{D_s^+}$ is much smaller than $A_{CP}^{D^0}$
because there is no mixing between $D_s^+$ and $D_s^-$.
We estimate the effect (0.01\%) due to the possible difference between
$A_{\rm{FB}}^{D_s^+}$ and $A_{\rm{FB}}^{D^0}$; we compare
$A_{\rm{FB}}$ in $D_s^+ \to \phi \pi^+$ and $D^+ \to \phi \pi^+$ decays.
In addition to binning, the following sources are considered for the uncertainty in
$A_{CP}^{\rm raw}$.
Based on the Monte Carlo simulation, we estimate that the peaking background
contributes 0.12\% of the signal.  The dominant contribution comes from
$B^+ \to \bar{D^0} \pi^+$ with $\bar{D^0} \to K^+\pi^-$ where the $\pi^+\pi^-$ tracks are
misidentified as $\ell^+\ell^-$, and from $B^+ \to J/\psi K^*(892)^+$.
We estimate the systematic error (0.01\%) using 
$A_{CP}(B^+ \to \bar{D^0} \pi^+) = 0.8 \pm 0.8\%$~\cite{BELLE-D0PIP} and 
$A_{CP}(B^+ \to J/\psi K^*(892)^+) = -4.8 \pm 3.3\%$~\cite{BABAR}.
The systematic error due to possible asymmetry in non-peaking background
(0.022\%) is estimated by repeating fits allowing the background asymmetry 
to vary.

The measurement and correction for kaon detection asymmetry is verified
by repeating the whole procedure with different requirements on
kaon identification.
  The corrected $A_{CP}$ values are stable 
within the statistical uncertainty and estimated systematic error.

Adding all systematic errors above in quadrature, the total systematic
error is estimated to be 0.22\%.

\begin{table}[htb]
\caption{
  Summary of systematic uncertainties.
}
\label{tbl:B_FITTING_RESULTS}
\scalebox{0.85}{
\begin{tabular}
 {@{\hspace{0.5cm}}c@{\hspace{0.5cm}}  @{\hspace{0.5cm}}c@{\hspace{0.5cm}} @{\hspace{0.5cm}}c@{\hspace{0.5cm}}}
    \hline
    \hline
                                     &    Source                                                &  $\%$      \\
    \hline  
    $A_{CP}^{\rm raw}$               &  Peaking background                                      & 0.01       \\
                                     &  ARGUS background                                        & 0.02      \\
                                     &  $M_{\rm bc}$ bin width                                  & $<$ 0.01      \\
                                     &  $(p_{\rm lab}^K,\cos\theta_{\rm {lab}}^{K})$ binning    & 0.02      \\
    \hline  
    $A_{\rm {rec}}^{D_s^+}$         &  $D_s^+ \to \phi\pi^+$ statistics                     & 0.17      \\
                                     &  $M_{\phi\pi^+}$ bin width                               & $<$ 0.01      \\
                                     &  $M_{\phi\pi^+}$ mass window                             & 0.02      \\
                                     &  ($p_{\rm {lab}}^{\pi^+}$,
                                         $\cos\theta_{\rm {CMS}}^{D_s^+}$,
                                         $\cos\theta_{\rm {lab}}^{\pi^+}$) binning            & 0.08      \\
                                    &  Empty bins                                           & $<$ 0.01      \\
                                     &  $\phi \to K^+K^-$ asymmetry                             & 0.05      \\
  \hline
    $A_{\rm {rec}}^{D^0}$            &  $D^0 \to K^- \pi^+$ statistics & 0.07       \\
                                     &  $M_{K^-\pi^+}$ bin width                                & $<$ 0.01      \\
                                     &  $M_{K^-\pi^+}$ mass window                              & $<$ 0.01      \\
                                     &  $(p_{\rm lab}^K,\cos\theta_{\rm {lab}}^{K})$ binning    & 0.04      \\
                                     &  possible $A_{CP}^{D^0}$                                 & 0.01       \\
                                     &  $A_{\rm{FB}}^{D_s^+} = A_{\rm{FB}}^{D^0}$ assumption    & 0.01      \\
    \hline
                              &      Total                                   & 0.22       \\
    \hline
    \hline
\end{tabular}}
\end{table}

In conclusion, using 772 $\times$ $10^6$ $B\overline{B}$ meson pairs,
we have measured the $CP$-violating charge asymmetry $A_{CP}(B^+ \to J/\psi K^+)$ to be ($-0.76 \pm 0.50 \pm 0.22$)\%,
where the first uncertainty is statistical and second is systematic.
No significant evidence of $CP$ violation
is observed. Our measurement is consistent with
the world average (0.9 $\pm$ 0.8)\%~\cite{PDG}.
This result significantly improves the precision from
previous measurements at
$e^+e^-$ collider experiments~\cite{CLEO, BABAR} and supersedes our earlier result~\cite{LAST-BELLE}.
The Belle result is slightly more precise than the D0 measurement~\cite{D0}.
The two most precise measurements to date differ by less than two standard deviations.

We thank the KEKB group for excellent operation of the
accelerator, the KEK cryogenics group for efficient solenoid
operations, and the KEK computer group and
the NII for valuable computing and SINET3 network support.  
We acknowledge support from MEXT, JSPS and Nagoya's TLPRC (Japan);
ARC and DIISR (Australia); NSFC (China); MSMT (Czechia);
DST (India); MEST, NRF, NSDC of KISTI, and WCU (Korea); MNiSW (Poland); 
MES and RFAAE (Russia); ARRS (Slovenia); SNSF (Switzerland); 
NSC and MOE (Taiwan); and DOE (USA).



\begin{thebibliography}{99}
\bibitem{CC}
Charge-conjugate states are implied throughout this paper unless stated otherwise.
\end{thebibliography}
\end{document}